\documentstyle[preprint,aps,prc,epsf,floats,tighten]{revtex} 
\begin{document}

\draft
\preprint{CEBAF-TH-95-03, DOE/ER/40762-055, UMPP95-091}

\title{ Instant Two-Body Equation in Breit Frame }

\author{ N. K. Devine }
\address{Continuous Electron Beam Accelerator Facility \\
12000 Jefferson Avenue, Newport News, VA 23606}

\author{ S. J. Wallace}
\address{ Department of Physics and Center for Theoretical Physics \\
          University of Maryland, College Park, MD 20742 }

\date{Jan 27, 1995}   

\maketitle

\begin{abstract}
A quasipotential formalism for elastic scattering from relativistic bound
states is based on applying an instant constraint
to both initial and final states in the Breit frame.
This formalism is advantageous for the analysis of electromagnetic
interactions because current conservation and four momentum conservation
are realized within a three-dimensional formalism.\cite{QPF}
Wave functions are required in a frame
where the total momentum is nonzero, which means that the
usual partial wave analysis is inapplicable.  In this work, the
three-dimensional equation is solved
numerically, taking into account the relevant
symmetries.  A dynamical boost of the interaction also is needed for
the instant formalism, which in general requires that the boosted
interaction be defined
as the solution of a four-dimensional equation.  For the case of a
scalar separable interaction, this equation is solved and the
Lorentz invariance of the three-dimensional formulation using the boosted
interaction is verified.  For
more realistic interactions, a simple approximation is used to
characterize the boost of the interaction.
\end{abstract}

\narrowtext


\section{INTRODUCTION}

The theory of relativistic bound states in quantum field theory
features four-dimensional equations and, in principle, an
infinite number of degrees of freedom.  Practical methods to
solve the problem are not available and it is therefore common to reduce
its complexity.  In this paper, we discuss a covariant reduction to
three dimensions and a finite number of degrees of freedom.\cite{QPF}
The interactions used are instantaneous.  The formalism is tractable and it
is applied to the deuteron bound state problem.

A general technique for reducing four-dimensional dynamics to three
dimensions is to introduce a constraint which fixes one
component of a four-vector
in terms of the others.\cite{BbSLT} This may be done covariantly, but
the resulting formalism may possess unphysical
asymmetries or singularities.  An example is when one particle is
constrained to its mass shell, and a second, identical particle is
not.\cite{GrossTwoBody82}
Symmetry with respect to exchange of particle labels is lost.
It is possible to respect the
Pauli Principle by use of appropriately defined interactions, but they are
cumbersome and they possess unphysical singularities which
must be removed by hand. \cite{Gross92}

Symmetrical three-dimensional
reductions have been used recently by Tjon and
collaborators.\cite{TjonQPReview,HummelMEC}  Our work is similar but is based
on the three-dimensional formalism developed by Mandelzweig and
Wallace\cite{ManWall87,WallaceMan89}, with an instant constraint at
non-zero total momentum.\cite{QPF}
An instant constraint maintains
symmetry with respect to exchange of particle labels and yields a
constrained equation for relativistic bound states with no
unphysical singularities.  In this regard, it is an attractive alternative
to reductions in which one particle is constrained to its mass shell.
Although unphysical singularities are avoided, they are not entirely
absent.  They are expected to appear in
corrections to the theory and in the four-dimensional equations for the boost
of the interaction.  Unphysical singularities are an unsolved problem
for quasipotential approaches.  In our approach to elastic
electron scattering from a two-body bound state, they play no role.

A theory of bound states
needs to be complemented by a
corresponding theory of currents, e.g., the electromagnetic current.
It is necessary to maintain a well-defined connection to quantum
field theory and, for this reason, the Bethe-Salpeter formalism is
the preferred starting point for the two-body problem.  The associated
currents have been formulated by Mandelstam in a celebrated
paper.\cite{Mandelstam55}  In
general, the currents depend on the interactions used to describe the
bound state and a consistent relationship
between the two is mandatory in
order to conserve the electromagnetic current.  An attractive feature of
the instant formalism we discuss is that a Ward-Takahashi identity is realized
at the level of the relativistic impulse approximation, thus guaranteeing
current conservation for elastic matrix elements.  There is a
price associated with the current conservation.  In the analysis of
elastic electromagnetic scattering, a boost of the interaction is necessary
in order to calculate
 the bound state
wave functions in the Breit frame, where the total three-momentum is
nonzero.

Dirac showed that the boost operator must depend on the interactions when
the generators of the Poincare group are quantized at an instant of
time.\cite{Dirac49}
This has its counterpart in the instant quasipotential formalism.
A dynamical, four-dimensional equation must be solved to determine the
instant quasipotential corresponding to different values of the total
three-momentum.
We distinguish between a kinematical boost and a dynamical one.
The former is simpler because it is effected in the same
fashion as for free particles and this is an advantage of a
covariant
formalism.\cite{GrossTwoBody82,Gross92}.
The latter is required in an instant formalism.  In the work of
Hummel and Tjon,\cite{HummelMEC} an instant quasipotential is
used together with a kinematical boost.  This
leads to an inconsistency with respect to current conservation
in matrix elements.  In the present work, we develop methods to
handle the dynamical boost that are consistent with current conservation
for electromagnetic interactions.

For the special case of a
scalar, separable interaction, the dynamical equation
corresponding to a boost of the interaction is solved exactly
and we verify that the mass of the deuteron is invariant when it is
used.  The boosted
interaction varies rather slowly with the momentum for the
deuteron and it may be approximated by a renormalization of the
strength of the
rest-frame interaction.  This approximation is used in our
analysis of electromagnetic form factors, which will be the subject of
another paper.

This work contains the analysis and methods of solution
for wave functions in the Breit frame.
Because the total three-momentum is
nonzero, the usual partial wave analysis is inapplicable.
The wave functions are obtained by solving the three-dimensional
integral equations using appropriately boosted interactions.
Section II reviews the quasipotential equations in a frame where
total three-momentum is nonzero.  Section III
discusses the equation which relates the quasipotential in
different frames and presents the solution for a scalar,
separable interaction.
In Section IV, we discuss the symmetries
of the relativistic bound state equations and use them to
reduce the equations to a solvable form for the case of a
boson-exchange interaction.  Section V presents
results for the deuteron wave functions and their variation with
the total momentum of the bound state.  Section VI presents
some concluding remarks.

\section{Relativistic bound-state equation}

The two-body equation with instant constraint has been
derived in Ref.~\cite{ManWall87} in the rest frame of the
two-body system.
The derivation incorporates crossed graphs using a form of the
eikonal approximation.  The same derivation carried out in a frame where
the total momentum is nonzero yields the
three-dimensional quasipotential equation,
\begin{equation}
\label{e:homoeq}
  \psi({\bf p};P) = g_{0}({\bf p};P)
  \int\!\!\frac{d^{3}k}{(2\pi)^{3}} \hat{V}({\bf p},{\bf k};P)
  \psi({\bf k};P) ,
\end{equation}
with relative and total momenta $p \equiv (p_{1}-p_{2})/2$ and
$P\equiv p_{1}+p_{2}$ ($p ^{0} = 0, {\bf{P}} \equiv P^{z}\hat{\bf z}$).
The three-dimensional propagator is
expressed in terms of projection operators for positive- and
negative-energy states as follows,
\begin{equation}
\label{e:g0QP}
    g_{0}({\bf p},P) = \sum_{\rho_{1},\rho_{2}} \frac{\Lambda_{1}^{\rho_{1}}
  ({\bf p}_{1}) \Lambda_{2}^{\rho_{2}}({\bf p}_{2}) }{
  (\rho_{1}+\rho_{2})(E _{P}/2) - \epsilon_{1} - \epsilon_{2}},
\end{equation}
where $\rho_{i}=\pm$, $\Lambda_{i}^{\rho_{i}}({\bf p}_{i}) =
\rho_{i} P _{i}  ^{\rho _{i}} \gamma^{0}_{i}$, where
$P _{i}  ^{\rho _{i}} \equiv
u^{\rho_{i}}_{i}(\rho_{i}{\bf p}_{i})
[u^{\rho_{i}}_{i}(\rho_{i}{\bf p}_{i})]^{\dagger}$
are projection operators obeying $P _{i} ^{\rho _{i}} P _{i}
^{\rho ' _{i}} = \delta _{\rho _{i}, \rho ' _{i}}.$
Dirac spinors used obey the hermitian
normalization condition~(\ref{e:hpwspinors}).
Moreover,
$\epsilon_{i} \equiv \sqrt{m^{2}+{\bf p}^{2}_{i}}$
(${\bf p}_{1,2}= {1\over 2}{\bf P}\pm {\bf p}$) and
$P^{0} = E _{P} \equiv \sqrt{M^{2}+{\bf P}^{2}}$
are on-shell energies, and nucleon and deuteron masses are $m$ and $M$.
This propagator differs from the Dirac two-body propagator of
Refs.~\cite{ManWall87} only by use of the
instant constraint in the frame where the two-body system moves with
momentum ${\bf P}$, instead of in the rest
frame.

As discussed in Ref.~\cite{WallaceMan89}, it is possible
to use the constraint $P \cdot
p = 0$ to develop a covariant formalism
corresponding to instant interactions in the rest frame.
If this is done, solutions of the equation can be
boosted kinematically on a 3D surface embedded in the 4D space and
defined by the constraint $p \cdot P = 0$.
However, this constraint is not compatible with the momentum
transfer $q$ in interactions.
If in the initial state the constraint $P \cdot p = 0$ is satisfied,
the corresponding constraint for the final state, namely
$(P +q)\cdot (p \pm {1 \over 2} q) = 0$, depending on which particle
absorbs the momentum transfer, cannot be satisfied.
Absorption of the momentum transfer requires that the wave function be
known off the 3D surface defined by the constraint.
In contrast, a compatible formalism can be obtained by adopting the
constraint $p ^{0} = 0$ in the Breit frame in place of
the covariant one.  The Breit frame corresponds to initial momentum
$P_{i} = (E_{P}, 0, 0, -{1 \over 2} q)$ and final momentum
$P_{f} = (E_{P}, 0, 0, {1 \over 2}q)$,
where $E_{P} = \sqrt{M ^{2} + q^{2}/4 }$.
It has the special property that $q ^{0} = 0$ for elastic
interactions and thus is consistent with conservation of the
four-momentum and three-dimensional wave functions.

The inverse propagator is easily obtained from the projection property
and it is,
\begin{equation}
\label{e:invprop}
  g^{-1}_{0}({\bf p};P) = \gamma^{0}_{1}\gamma^{0}_{2} \left[
    (E_{P}/2 - \hat{\rho}_{1}\epsilon_{1}) \hat{\rho}_{2} +
    (E_{P}/2 - \hat{\rho}_{2}\epsilon_{2}) \hat{\rho}_{1} \right] ,
\end{equation}
with $\hat{\rho}_{i}=h({\bf p}_{i})/\epsilon_{i}$
(thus $\hat{\rho}_{i} u^{\pm}_{i}({\pm\bf p}_{i})
 = \pm u^{\pm}_{i}({\pm\bf p}_{i})$).
The normalization condition for the two-body wave function is,
\begin{eqnarray}
\label{e:psiNorm}
   1 &=& {1\over {2 E _{P}}}
 \int\!\!\frac{d^{3}p d^{3}k}{(2\pi)^{6}} \overline{\psi}({\bf p};P) \left[
 \gamma^{0}_{1}\gamma^{0}_{2} \frac{\hat{\rho}_{1}+\hat{\rho}_{2}}{2}
 (2\pi)^{3}\delta({\bf p}-{\bf k}) \right. 
   \left. - \left( \frac{\partial}{\partial P^{0}}
     \hat{V}({\bf p},{\bf k},P) \right) \right] \psi({\bf k};P) .
\end{eqnarray}

\section{Boost of the Quasipotential}

In quasipotential approaches, the quasipotential kernel is formally
related to the Bethe-Salpeter kernel by,
\begin{equation}
\label{e:K_QP}
  K^{QP}(P) = K^{BS}(P) + i K^{BS}(P) \Delta(P)  K^{QP}(P) ,
\end{equation}
where a four-dimensional integration is implied over relative momentum, $p$,
and
\begin{equation}
\Delta (P) \equiv G ^{BS}_{0} (P) - G ^{QP}_{0} (P) ,
\end{equation}
\begin{eqnarray}
G ^{BS} (P) &=&
  \frac{ \bigl[ p_{1}\cdot \gamma_{1} + m \bigr] }
       { \bigl[ p_{1}^{2} - m^{2} + i \eta \bigr] } \,
  \frac{ \bigl[ p_{2}\cdot \gamma_{2} + m \bigr] }
       { \bigl[ p_{2}^{2} - m^{2} + i \eta \bigr] } ,
\end{eqnarray}
and
\begin{equation}
\label{e:GQP}
G^{QP}_{0}(P) = -i g_{0}({\bf p},P) 2\pi \delta(p^{0})
\end{equation}
in our approach.  The quasipotential $\hat{V}$ used in
Eq.~(\ref{e:homoeq}) corresponds to $K ^{QP}$ with initial and
final momenta restricted to the constraint space, i.e., $p _{0}=0$.
As emphasized in the notation, the quasipotential
depends in general on the total four-momentum, $P$.
The quasipotential propagator involving $p ^{0} = 0$
corresponds to different constraints for different values of $P$.
Only when the constraint is expressed covariantly does it
have the same physical meaning at all $P$ values.

In order to obtain the quasipotential corresponding
to an instant constraint at different four-momenta,
one must in general solve Eq.~(\ref{e:K_QP}) at each value of $P$.
Alternatively, it is possible to eliminate the
Bethe-Salpeter kernel to arrive at a direct relation of the
quasipotential corresponding to two different momenta $P$ and $P_{0}$.
We do this for a somewhat simplified case where the Bethe-Salpeter kernel
is assumed not to depend on the total momentum, $P$, and find
\begin{eqnarray}
\label{e:QPchange}
K^{Q}(P) &=&
  K^{QP}(P _{0}) + i K^{QP}(P_{0}) \: [\Delta (P) - \Delta (P_{0}) ] \:
        K^{QP}(P) .
\end{eqnarray}
In the present work, we consider the boost of the
quasipotential from the rest frame, where momentum is $P_{0} = (M, {\bf 0})$,
to a frame where the four momentum is $P = (E _{P}, {\bf P})$.

The chief complication in solving for the quasipotential lies in the
implied 4D integration.  However, the problem is soluble for the case of a
separable Bethe-Salpeter kernel of the form,
\begin{equation}
K  =  |\chi\rangle k \langle\chi| ,
\end{equation}
where $ |\chi\rangle $ carries the dependence on relative momentum,
$p$, and $k$ is a $16 \times 16$ matrix.   For the discussion
of the separable potential case, we omit
parts of the quasipotential propagator which arise from the
treatment of crossed Feynman graphs because these contributions are not
meaningful for a separable potential.  A simpler
quasipotential propagator is used which is expressible as,
\begin{equation}
G ^{QP} = 2 \pi \delta (p ^{0}) \int \frac{dp^{0}}{2 \pi} G^{BS}  ,
\end{equation}
which agrees with Eqs.~(\ref{e:GQP}) and (\ref{e:g0QP}) in the
$+\,+$ and $-\,-$ rhospin states, but is zero in the $+\,-$ and $-\,+$
rhospin states.

It follows that the quasipotential also takes a separable form
\begin{equation}
K ^{QP} (P ) =  |\chi\rangle k ^{QP} (P) \langle\chi| .
\end{equation}
where $k ^{QP}(P)$ is also a $16 \times 16$ matrix in the two-particle
Dirac space.  It is related to the Bethe-Salpeter kernel by the matrix
equation,
 \begin{equation}
\label{e:kQP}
k ^{QP} (P) = k  + k  \overline{\Delta} (P)   k ^{QP} (P) ,
 \end{equation}
where
\begin{equation}
\overline{\Delta} (P) = i \int \frac{d ^{4} p}{(2 \pi )^{4}} \chi(p)
^{2} \Delta (P)
\end{equation}
is a $16 \times 16$ matrix.
Alternatively, one may relate the quasipotential matrices at two
different momenta by use of Eq.~(\ref{e:QPchange}), which leads
to the matrix equation,
\begin{equation}
\label{e:boost}
k ^{QP} (P) = k ^{QP} (P_{0}) + k ^{QP} (P_{0}) [ \overline{\Delta} (P) -
\overline{\Delta} (P_{0}) ] k ^{QP} (P) .
\end{equation}

Either Eq.~(\ref{e:kQP}) or (\ref{e:boost}) is readily solved,
e.g., Eq.~(\ref{e:kQP}) yields
\begin{equation}
\label{e:matrixkQP}
k ^{QP} (P) = \bigl\{ \openone - k \overline{\Delta} (P) ] \bigr\}
^{-1} k .
\end{equation}

\subsection{Analysis for a scalar, separable potential }

In order to gain insight into the nature of the boost of the
quasipotential, we have solved for $k^{QP}(P)$ for the case of a deuteron
bound by a scalar, separable interaction.  This means that $k$ is
a coupling constant, $g$, times
the direct product of unit matrices in each particle's Dirac
space.  Taking symmetries into account, we find
\begin{eqnarray}
\overline{\Delta} (P) &=&
    \bigl( {1 \over 2} P \cdot \gamma_{1} + m \bigr)
    \bigl( {1 \over 2} P \cdot \gamma_{2} + m \bigr) \Delta _{1}(P)
   - \frac{P\cdot\gamma_{1}}{M} \frac{P\cdot\gamma_{2}}{M} \Delta _{2}(P)
   - \frac{\widetilde{P}  \cdot \gamma_{1}}{M}
       \frac{\widetilde{P}  \cdot \gamma_{2}}{M} \Delta _{3}(P) \nonumber\\
  && - \biggl( \frac{P\cdot\gamma_{1}}{M}
\frac{\widetilde{P}\cdot\gamma_{2}}{M}
             + \frac{\widetilde{P}\cdot\gamma_{1}}{M}
               \frac{ P\cdot\gamma_{2}}{M}  \biggr) \Delta _{4}(P) 
       + \gamma _{1 \perp} \cdot \gamma _{2 \perp} \Delta _{5} (P) ,
\end{eqnarray}
where $P = (E_{P}, 0, 0, P)$, $\widetilde{P} = (P, 0, 0, E_{P})$ and thus
$P \cdot \widetilde{P} = 0$.   Vectors with the $\perp$ subscript have only
x- and y-components and they are orthogonal to $P$ and
$\widetilde{P}$.  The quantities $\Delta _{n}(P)$ for n = 1 to 5
in this expression
are determined by the following equations,
\begin{equation}
\label{e:deltan}
\Delta _{n}(P) = A^{BS} _{n} (P) - A^{QP} _{n} (P),
\end{equation}
where
\begin{eqnarray}
\label{e:AnBS}
A^{BS}_{n}(P) &=& i \int \frac{d^{4}p}{(2\pi )^{4}}
    \left[\chi(p)\right]^{2} 
   \bigl[ p_{1}^{2} -m^{2} +i\eta \bigr]^{-1}
           \bigl[ p_{2}^{2} -m^{2} +i\eta \bigr]^{-1} O_{n},
\end{eqnarray}
\begin{eqnarray}
\label{e:AnQP}
A^{QP}_{n}(P) &=& i \int \frac{d^{3}p}{(2\pi )^{3}}
  \left[\chi({\bf p})\right]^{2} \int \frac{dp^{0}}{2 \pi}
    \bigl[ p_{1}^{2} -m^{2} +i\eta \bigr]^{-1}
           \bigl[ p_{2}^{2} -m^{2} +i\eta \bigr]^{-1} O_{n},
\end{eqnarray}
and
\begin{eqnarray}
O _{1} &=& 1,\\
O _{2} &=& \biggl(\frac{p \cdot P}{M}\biggr)^{2}, \\
O_{3} &=& \biggl(\frac{p \cdot \widetilde P}{M}\biggr)^{2}, \\
O_{4} &=& \biggl(\frac{p \cdot P}{M}\frac{p \cdot \widetilde P}{M}\biggr), \\
O_{5} &=& p _{x} ^{2} .
\end{eqnarray}
Symmetries in the Bethe-Salpeter case cause $A _{3} ^{BS} =  A _{5}
^{BS}$, and $A _{4} ^{BS} =0 $.

It is convenient to take matrix elements of $\overline{\Delta} (P)$
between plane-wave Dirac spinors depending on the total
momentum and defined as follows,
\begin{equation}
u ^{+} ({\bf{P}}) = N _{P} \pmatrix{ 1 \cr \frac{\textstyle{{\bbox{\sigma}}
 \cdot {\bf{P}}}}{\textstyle{E _{P} + M}} \cr} ,
\label{uplus}
\end{equation}
\begin{equation}
u ^{-} ({\bf{P}}) = N _{P} \pmatrix{ \frac{{\textstyle{\bbox{\sigma}} \cdot
{\bf{P}}}}{\textstyle{E _{P} + M}} \cr 1 \cr}.
\label{uminus}
\end{equation}
These states are eigenfunctions of $\gamma \cdot P /M = \pm 1$
and the
normalization factor is
\begin{equation}
N _{P} \equiv \sqrt{\frac{E _{P} + M}{2 M }} ,
\label{norm}
\end{equation}
where $E _{P} = \sqrt{P ^{2} + M ^{2}}$.  Negative
energy states have negative norm: $\bar{u} ^{-} (P) u^{-}(P) = -1$.
The matrix form of $\overline{\Delta} (P)$ is,
\widetext
\begin{eqnarray}
[\overline{\Delta}(P)] =
\pmatrix{ \overline{\Delta}^{\,++,++} (P) & \overline{\Delta}_{\sigma}(P)
  & \sigma _{2z} \Delta _{4}(P)  & \sigma _{1z} \Delta _{4}(P) \cr
\overline{\Delta} _{\sigma}(P) &  \overline{\Delta} ^{\,--,--}(P)
  &  \sigma_{1z} \Delta _{4}(P) & \sigma _{2z} \Delta _{4}(P) \cr
\sigma_{2z} \Delta _{4}(P) & \sigma _{1z} \Delta _{4}(P)
  & \overline{\Delta} ^{\,+-,+-} (P) & \overline{\Delta} _{\sigma}(P) \cr
\sigma_{1z} \Delta _{4}(P) & \sigma _{2z} \Delta _{4}(P)
  & \overline{\Delta}_{\sigma}(P)  & \overline{\Delta}^{\,-+,-+} (P) \cr } ,
\end{eqnarray}
\narrowtext
where
\begin{eqnarray}
\label{e:Delta++,++}
 \overline{\Delta} ^{\,++,++} (P) &=& \overline{u ^{+}_{1} } ({\bf P})
   \overline{u ^{+}_{2} } ({\bf P}) \overline{\Delta} (P) u^{+}_{1}({\bf P})
   u ^{+}_{2}  ({\bf P}) 
 = ({1 \over 2} M + m ) ^{2} \Delta _{1}(P) - \Delta _{2}(P) ,
\end{eqnarray}
\begin{eqnarray}
 \overline{\Delta} ^{\,--,--} (P)  &=& \overline{u ^{-}_{1} } ({\bf P})
   \overline{u ^{-}_{2} } ({\bf P}) \overline{\Delta} (P) u ^{-}_{1}({\bf P})
   u ^{-}_{2}  ({\bf P}) 
 = ({1 \over 2} M - m ) ^{2} \Delta _{1}(P) - \Delta _{2}(P) ,
\end{eqnarray}
\begin{eqnarray}
 \overline{\Delta} ^{\,+-,+-} (P) &=& \overline{u ^{+}_{1} } ({\bf P})
   \overline{u ^{-}_{2} } ({\bf P}) \overline{\Delta} (P) u ^{+}_{1}({\bf P})
   u ^{-}_{2}  ({\bf P}) 
 = ({1 \over 4}  M^{2} - m^{2} ) \Delta _{1}(P) - \Delta _{2}(P) ,
\end{eqnarray}
\begin{eqnarray}
 \overline{\Delta} ^{\,-+,-+} (P) &=& \overline{u ^{-}_{1} } ({\bf P})
   \overline{u ^{+}_{2} } ({\bf P}) \overline{\Delta} (P) u ^{-}_{1} ({\bf P})
   u ^{+}_{2}  ({\bf P}) 
 = ({1 \over 4}  M^{2} - m^{2} ) \Delta _{1}(P) - \Delta _{2}(P) ,
\end{eqnarray}
and
\begin{equation}
 \overline{\Delta} _{\sigma} = -\sigma _{1z} \sigma _{2z} \Delta _{3}(P) +
   \bbox{\sigma}_{1 \perp} \cdot \bbox{\sigma}_{2\perp } \Delta _{5}(P) .
\end{equation}

It follows that the solution of Eq.~(\ref{e:matrixkQP}) may be
found and thus the full structure of the quasipotential
displayed: $k^{QP} = \{\openone - g \overline{\Delta} \}^{-1} g\openone$
leads to the matrix equation
$[\openone - g \overline{\Delta}] \, [\openone] \, [k^{QP}] = g [\openone]$,
thus,
\begin{eqnarray}
\label{e:kQP-1}
  [k^{QP}] = \bigl( \, g^{-1} \, [\openone]
    - [\openone] \, [\overline{\Delta}] \, [\openone] \, \bigr)^{-1} .
\end{eqnarray}
A subtlety here is that the matrix for the unit operator is $[\openone] =
diag \{ 1, 1, -1, -1\}$.
It is straightforward to realize $k ^{QP}$ numerically by calculating the
inverse implied in Eq.~(\ref{e:kQP-1}).

In the case of a separable potential, one may readily solve the
Bethe-Salpeter wave equation or the quasipotential wave equation, using
kernels appropriate to each.  For the Bethe-Salpeter equation,
we find
\begin{equation}
\label{e:BSsep}
\phi ^{BS} (P) = i \langle\chi| G ^{BS} (P) |\chi\rangle  \cdot k \cdot \phi
^{BS},
\end{equation}
and for the quasipotential equation,
\begin{equation}
\label{e:QPsep}
\phi ^{QP} (P) = i \langle\chi| G ^{QP} (P) |\chi\rangle  \cdot
  k^{QP} (P) \cdot \phi^{QP}.
\end{equation}
These are equivalent if $k ^{QP}$ is obtained from Eq.~(\ref{e:kQP}).
The quantity $i \langle\chi| G ^{BS} (P) |\chi\rangle$ has the same form as
$\overline{\Delta}(P)$ using the $A ^{BS} _{n}$ in place of $\Delta_{n}(P)$.
Similarly, $i \langle\chi| G ^{QP} (P) |\chi\rangle $
has the same form using  $A ^{QP} _{n}$ in place of $\Delta _{n}(P)$.

Numerical calculations for the scalar separable potential have
been performed based on
using $\chi (p) = ( 1 + p ^{2} / \mu ^{2} ) ^{-2}$
with $\mu =$ 200 MeV for the separable potential.
The coupling constant $g$ is determined by the condition that a bound state
exists for $M$ equal to the deuteron mass in the Bethe-Salpeter equation.
Solutions corresponding to differing values of ${\bf P}$ are obtained from
both equations and they demonstrate that
the deuteron mass is invariant when the boosted
quasipotential is used in Eq.~(\ref{e:QPsep}).

   Owing to dominance of the $+\,+$ states in the case of weak
binding, an approximate characterization of the boost of the
quasipotential is possible.  A significant part of the effect is to renormalize
the positive-energy matrix element in comparison to its
value in the rest frame.   The ratio of matrix elements defined as in
Eq.~(\ref{e:Delta++,++}) and calculated in the rest
and moving frames is,
\begin{equation}
\label{e:lambda1}
\lambda ({\bf P}) = \frac{ {1 \over 4} Tr\left[ k^{QP}(P_{0})
 \right]^{++,++} }{  {1 \over 4} Tr \left[ k^{QP}(P) \right] ^{++,++} },
\end{equation}
where the trace is over spins.  This ratio
may be used to approximate the boost
as a similar renormalization of all matrix elements,
\begin{equation}
\label{e:lambdaboost}
 k^{QP}(P) \approx k^{QP}(P _{0}) / \lambda ({\bf P}).
\end{equation}

\begin{figure}
  \centerline{\epsfxsize=8cm  \epsffile{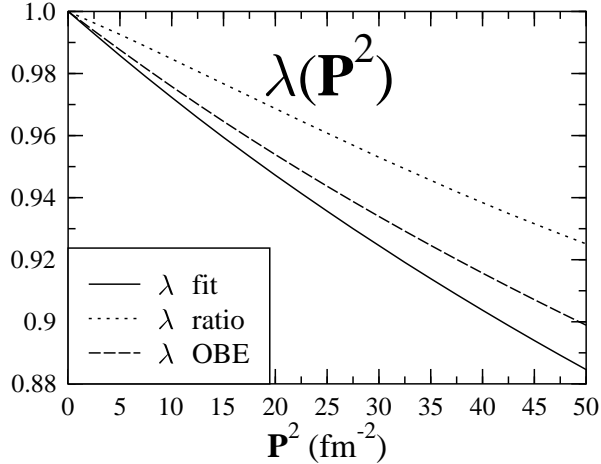} }
\caption{
Renormalization factor of scalar quasipotential.
Solid line shows $\lambda $ for invariant deuteron
mass and dotted line shows prediction of Eq.~(\protect\ref{e:lambda1}).
Dashed line shows
result for a scalar, meson-exchange interaction.
\label{f:lam_sep}  }
\end{figure}

A ``fit'' renormalization parameter can be determined as a function
of ${\bf P}$ by using Eq.~(\ref{e:lambdaboost}) and solving
Eq.~(\ref{e:QPsep}) for the value of $\lambda({\bf P})$ such that the
mass is invariant.
Figure~\ref{f:lam_sep} shows the variation of $\lambda({\bf P})$
with momentum for
this fit case.  For comparison, we show the ``ratio'' prediction based on
Eq.~(\ref{e:lambda1}), which is in reasonable agreement with the
``fit'' value.
Also shown in Figure~\ref{f:lam_sep} is $\lambda({\bf P})$
fit to yield an invariant deuteron mass with
the propagator of the next section,
but using only the (scalar) $\sigma$ meson.
The fit renormalization parameter $\lambda({\bf P})$ decreases for
increasing ${\bf P}^{2}$ in a qualitatively similar
fashion for scalar potentials of either the
separable or one-boson-exchange type.
Equation (\ref{e:kQP-1}) would remain diagonal and proportional to
$diag \{ 1, 1, -1,-1 \}$ if the Dirac structure remained scalar in the boost.
As $|{\bf P}|$ increases, the matrix can be seen to change its structure.
Owing to the dominance of the $++,++$ matrix elements,
a significant part of the effect is a renormalization of the interaction.

If one has knowledge of the rest frame interaction $ k ^{QP} (P _{0})$,
Eq.~(\ref{e:boost}) may be used to determine $k ^{QP} (P)$.
This case is interesting because the NN interaction
is usually regarded as known in the rest frame and the problem is to
boost it to other frames.
A more accurate approximation than a simple renormalization of all
matrix elements is to expand perturbatively as follows,
\begin{equation}
\label{e:boost_pert}
k ^{Q}(P) \approx k ^{QP} (P _{0}) + k ^{QP} (P_{0}) \left[
\overline{\Delta} (P) - \overline{\Delta} (P _{0}) \right] k ^{QP} (P
_{0})  .
\end{equation}
Keeping the second order term in Eq.~(\ref{e:boost_pert}) produces
a form for $k ^{QP} (P)$ that eliminates most of the momentum dependence
of the mass.  If a renormalization factor $\lambda ({\bf P})$ is
used with the approximation of Eq.~(\ref{e:boost_pert}), then
a value $\lambda$ = 0.984 at ${\bf
P}^{2}$ = 50 fm$^{-2}$ is required to keep the mass invariant, as compared
with a value $\lambda$ = 0.882 when $k ^{QP} (P) \approx k ^{QP} (P_{0})
/ \lambda$ is used and $\lambda \equiv 1$ when the exact $k ^{QP} (P)$ is
used.

\subsection{Effective boost approximation }

For the one-boson-exchange potential (in the rest frame) we use the Bonn
potential (Bonn B, energy-independent, Thompson
propagator)\cite{DevineThesis,BonnABC}.
This potential includes scalar, pseudo-vector,
and vector meson exchanges ($\sigma,\delta,\eta,\pi,\omega,\rho$)
and is detailed in Appendix~\ref{app:Vpartial}.
When projected onto positive-energy plane-waves ($\rho_{1}\rho_{2}=++$)
in the center-of-mass frame, the Dirac two-body $g_{0}$ of
Eq.~(\ref{e:g0QP}) reduces to the
Thompson propagator and $V$ reduces to the Bonn potential.
When negative-energies are included, there arise couplings
in the quasipotential for which initial state rho spins of both
particles are
opposite to the final state rho spins.  For example, $V
^{++,--}$.  An analysis of
Feynman diagrams where such couplings arise shows that they
generally are suppressed strongly in comparison with other
couplings owing to the necessity of large time-like momentum
transfer of order $2m $.  For all other couplings, the
instant constraint provides a reasonable starting approximation.
To accommodate this fact, we eliminate the suppressed couplings
by setting them to zero, i.e.,
$[\gamma^{0}_{1}\gamma^{0}_{2}V]^{\rho'_{1},\rho'_{2},\rho_{1},\rho_{2}}
\to 0$ when $\rho'_{1}=-\rho_{1}$ and $\rho'_{2}=-\rho_{2}$.

To obtain the correct deuteron binding energy
when negative energy sectors are
included, we modify the Bonn B potential by increasing the scalar
attraction about $6 \%$, from $g^{2}_{\sigma}/4\pi=8.0769$ to $8.5503$.
(See Table~\ref{t:mesonparams}.)

\begin{figure}
  \centerline{\epsfxsize=8cm  \epsffile{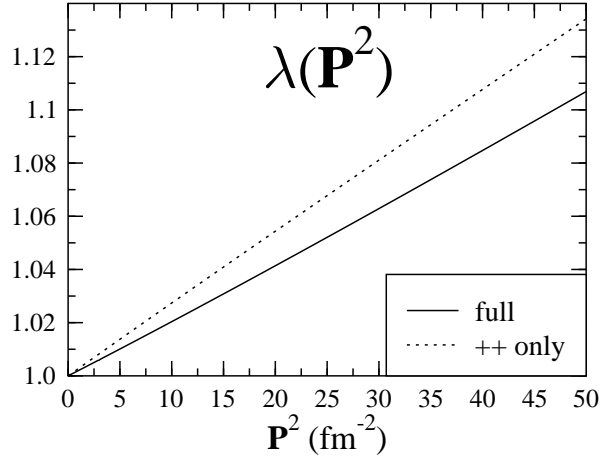} }
\caption{The scaling of the potential, $\hat{V}(p'-p,{\bf P})
  = \hat{V}(p'-p) / \lambda ({\bf P}^{2})$,
  that produces constant deuteron mass,
  $M= (E_{P}^{2}-{\bf P}^{2})^{1/2} = 2m-2.22464MeV$: full
  propagator (solid), $++$ states only (dotted).
\label{f:lambdaVSqsq} }
\end{figure}

For the meson-exchange interaction,
the simple approximation discussed above has been used to boost
the quasipotential:
$\hat{V}({\bf p}'-{\bf p},{\bf P})
= \hat{V}({\bf p}'-{\bf p}) / \lambda ({\bf P})$,
where $\lambda ({\bf P})$ is fit to produce the correct deuteron total
energy, $E_{P}= (M^{2}+{\bf P}^{2})^{1/2}$, using the two-body equation
and propagator of Eqs.~(\ref{e:homoeq}) and~(\ref{e:g0QP}).
This approximation can be shown to be consistent with current
conservation when used in the analysis of elastic form factors.
Note that the Dirac spinors appropriate to a moving deuteron
are treated exactly and that the factor $\lambda ({\bf P})$
approximates only the additional change of the potential
required in an instant formalism.
Figure~\ref{f:lambdaVSqsq} shows that the required change of the
potential is modest, with $\lambda$ varying linearly vs.\
${\bf P}^{2}$ over a wide range of values.
For the full meson-exchange interaction, the renormalization
factor $\lambda({\bf P}^{2})$ increases with ${\bf P}^{2}$,
in contrast with the result of Figure \ref{f:lam_sep} for a scalar interaction.
This is caused by the differing boost factors required for
different types of interaction.
When $\lambda ({\bf P})=1$ is used, the potential is too
attractive and the binding energy of the deuteron increases from
$2m - M \approx 2.2MeV$ at ${\bf P}^{2}= 0$ to
$2m - M \approx 4.2MeV$ at ${\bf P}^{2}=50fm^{-2}$.
We have calculated the deuteron form factors based on
$\lambda ({\bf P})=1$ for comparison with those based on $\lambda
({\bf P})$ determined so that $M$ is invariant.  The differences are
small.

\section{Symmetries and reduction of the equation}

The homogeneous equation is symmetric with respect to
the operations of spatial reflection (${\cal P}$),
particle exchange (${\cal P} _{\!\!12}$), and time-reversal
(${\cal T}$),
i.e., $g_{0} = {\cal P} g_{0} {\cal P}^{-1} =
 {\cal P}_{\!\!12} g_{0} {\cal P}_{\!\!12}^{-1} =
 {\cal T} g_{0} {\cal T}^{-1}$, and $V$ is assumed to behave similarly.
Thus it is possible to have
solutions of the homogeneous equation with good corresponding parities:
\begin{eqnarray}
  \psi^{M}_{J}({\bf p},{\bf P}) &=& \eta {\cal P}
  \psi^{M}_{J}({\bf p},{\bf P}) , \\
  \psi^{M}_{J}({\bf p},{\bf P}) &=& \eta_{12} {\cal P}_{\!\!12}
  \psi^{M}_{J}({\bf p},{\bf P}) , \\
  \psi^{M}_{J}({\bf p},{\bf P}) &=& \eta_{T} (-1)^{M} {\cal T}
  \psi^{-M}_{J}({\bf p},{\bf P}) ,
\end{eqnarray}
where the parity, exchange parity, and anti-linear time-reversal
operators are,
\begin{eqnarray}
  {\cal P} &=& \gamma_{1}^{0}\gamma_{2}^{0}  {\cal P}_{\!\!s}  ,
                                                  \label{e:parityop}\\
  {\cal P}_{\!\!12} &=& {\cal P}_{\!\!ex} \Pi , \label{e:exchparityop}\\
  {\cal T} &=& \sigma^{2}_{1}\sigma^{2}_{2} {\cal K} ,\label{e:timeparityop}
\end{eqnarray}
where ${\cal K}$ is the operator of complex conjugation and
${\cal K}$ and ${\cal P}_{\!\!s}$ reverse three-momenta
(${\cal P}_{\!\!s}{\bf p}= -{\bf p}$, ${\cal P}_{\!\!s}{\bf P}=-{\bf P}$);
$\Pi$ reverses relative four-momentum ($\Pi p = -p$);
and ${\cal P}_{\!\!ex}$ exchanges Dirac indices
(${\cal P}_{\!\!ex} \gamma_{2}^{\mu} = \gamma_{1}^{\mu} {\cal P}_{\!\!ex}$,
${\cal P}_{\!\!ex} u^{\rho_{1}}_{1}(\rho_{1} {\bf p}_{1}) =
u^{\rho_{1}}_{2}(\rho_{1} {\bf p}_{1}) {\cal P}_{\!\!ex}$, etc).
For the deuteron $\eta = \eta_{12} = \eta_{T} = +1$.
Note that ${\cal P} {\cal P}_{\!\!12}$ is equivalent to the H-parity
operator of Kubis\cite{Kubis}.

To find wave functions with good exchange parity, we rewrite
the homogeneous equation 
using $\psi(-{\bf p},{\bf P}) =
\eta_{12} {\cal P}_{\!\!ex} \psi({\bf p},{\bf P})$ and find that,
\begin{eqnarray}
\label{e:12homoeq}
   \psi({\bf p}';P) &=& g_{0}({\bf p}';P) \gamma^{0}_{1}\gamma^{0}_{2}
      \int_{p_{z}>0}\!\frac{d^{3}p}{(2\pi)^{3}}
      \gamma^{0}_{1}\gamma^{0}_{2}\left[
      \hat{V}({\bf p}',{\bf p};P) \right. 
     +\left. \hat{V}({\bf p}',-{\bf p};P) \eta_{12} {\cal P}_{\!\!ex} \right]
      \psi({\bf p};P) .
\end{eqnarray}
Note that the necessary range of integration of $p_{z}$ is halved.
To find wave functions with ${M}_{J}=0$ and good combined
${\cal P} {\cal T}$ parity,
we will form eight basis functions with $+{\cal P} {\cal T}$ parity,
and eight with $-{\cal P} {\cal T}$ parity.

%

    The Breit frame total angular momentum operator is,
${\bf J} = {\bf J}_{1} + {\bf J}_{2} = \vec{{\cal L}} + {\bf S}$,
where $\vec{{\cal L}} = {\bf l} + {\bf L}$,
${\bf l}={\bf r} \times {\bf p}$, ${\bf L}= {\bf R} \times {\bf P}$, and
${\bf S} = \frac{1}{2} (\bbox{\sigma}_{1}+\bbox{\sigma}_{2})$.
Because ${\bf J}^{2}$ and $J^{z}$ commute with $g^{-1}_{0}$ and $V$,
solutions of the homogeneous equation are eigenfunctions
of ${\bf J}^{2}$ and $J^{z}$.
Also, wave functions with polarization states $M_{J}=\pm 1$ can be obtained
from the $M_{J}=0$ state using raising and lowering operators,
$\psi^{M_{J}\pm 1} = \sqrt{(J+M_{J})(J-M_{J}+1)} J^{\pm} \psi^{M_{J}}$.
However, ${\bf l}$, ${\bf L}$,  and ${\bf S}$ do not separately commute with
$g^{-1}_{0}$, and  the usual $LSJ$ partial-wave analysis is inapplicable.
To proceed, we define cylindrical eigenfunctions
of $J^{z}$ which form a basis for the $\phi$ dependence
of the wave function:
\begin{equation}
    {\cal Y}^{M_{J}}_{s_{1},s_{2}}(\phi) =
  e^{i(M_{J}-s_{1}-s_{2})\phi} | s_{1} \rangle | s_{2} \rangle ,
\end{equation}
where ${\bf p} = (p^{xy} \cos(\phi), p^{xy} \cos(\phi), p^{z})$ is the
relative three-momentum,
$J^{z} {\cal Y}^{M_{J}} = M_{J} {\cal Y}^{M_{J}}$, and
$s_{i}=\pm \frac{1}{2}$ are the $+\hat{z}$ components of spin (parallel to
${\bf P} = P^{z} \hat{z}$).
A related set of eigenfunctions which we call the $PT$ basis
is given by,
\begin{eqnarray}
\label{e:Yofphi}
  {\cal Y}^{M_{J}}_{e1} &\equiv& \frac{1}{\sqrt{2}}
     ( {\cal Y}^{M_{J}}_{++} + {\cal Y}^{M_{J}}_{--} ) , \nonumber\\
  {\cal Y}^{M_{J}}_{o1} &\equiv& \frac{1}{\sqrt{2}}
     ( {\cal Y}^{M_{J}}_{++} - {\cal Y}^{M_{J}}_{--} ) , \nonumber\\
  {\cal Y}^{M_{J}}_{e0} &\equiv& \frac{1}{\sqrt{2}}
     ( {\cal Y}^{M_{J}}_{+-} + {\cal Y}^{M_{J}}_{-+} ) , \nonumber\\
  {\cal Y}^{M_{J}}_{o0} &\equiv& \frac{1}{\sqrt{2}}
     ( {\cal Y}^{M_{J}}_{+-} - {\cal Y}^{M_{J}}_{-+} ) ,
\end{eqnarray}
where the $\pm$ subscripts are shorthand for $\pm\frac{1}{2}, \pm\frac{1}{2}$.
It will be shown in Appendix~\ref{app:spinors} that these eigenfunctions
have good ${\cal P} {\cal T}$ parity if $M_{J}=0$.
Each set of eigenfunctions is orthonormal,
\begin{equation}
  \int \frac{d\phi}{2\pi} \left[ {\cal Y}_{a'}(\phi) \right]^{\dagger}
    {\cal Y}_{a}(\phi) = \delta_{a',a} ,
\end{equation}
where $a\equiv(s_{1},s_{2},M_{J})$ (or $a\equiv(\alpha,M_{J})$ with
$\alpha \in \{e1,o1,e0,o0\}$) and $a'$ is similarly defined.

To form a complete set of two-particle basis functions,
the angular eigenfunctions are combined with Dirac spinors
obeying the hermitian normalization, $u ^{\rho \dag}(\rho {\bf
p}) u ^{\rho '} (\rho ' {\bf p})
= \delta _{\rho \rho '}$
(\ref{e:hpwspinors}),
\begin{equation}
\label{pwbasis}
  \chi^{\rho_{1},\rho_{2}}_{a}({\bf p},{\bf P}) \equiv
 u^{\rho_{1}}_{1}(\rho_{1} {\bf p}_{1}) u^{\rho_{2}}_{2}(\rho_{2} {\bf p}_{2})
 {\cal Y}_{a}(\phi) .
\end{equation}
Either set of sixteen basis functions is an orthonormal set,
\begin{equation}
  \int \frac{d\phi}{2\pi} \left[
    \chi^{\rho'_{1},\rho'_{2}}_{a'}({\bf p},{\bf P}) \right]^{\dagger}
    \chi^{\rho_{1},\rho_{2}}_{a}({\bf p},{\bf P})
  = \delta_{\rho'_{1},\rho_{1}} \delta_{\rho'_{2},\rho_{2}}
    \delta_{a',a} ,
\end{equation}
and the wave functions are expanded in either set as follows,
\begin{equation}
     \psi^{M}_{J}({\bf p};{\bf P}) = \sum_{\rho_{1},\rho_{2},a}
          \chi^{\rho_{1},\rho_{2}}_{a}({\bf p},{\bf P})
          \psi^{\rho_{1},\rho_{2}}_{a}(p^{z},p^{xy};{\bf P}) .
\end{equation}

Using these plane-wave basis functions, the homogeneous
equation~(\ref{e:homoeq} or \ref{e:12homoeq}) can be written in
component form as
\begin{eqnarray}
\label{e:homoeq_comp}
  && \psi^{\rho'_{1},\rho'_{2}}_{a'}(p^{z},p^{xy};{\bf P}) =
    \int\! \frac{k^{xy}dk^{xy}k^{z}}{(2\pi)^{2}}
    G^{\rho'_{1},\rho'_{2}}_{a'}(p^{z},p^{xy};{\bf P})  
  V^{\rho'_{1},\rho'_{2}; \rho_{1},\rho_{2}}
    _{a';a}(p^{z},p^{xy};k^{z},k^{xy})
     \psi^{\rho_{1},\rho_{2}}_{a}(k^{z},k^{xy};{\bf P}) \nonumber \\
\end{eqnarray}
or
\begin{eqnarray}
\label{e:12homoeq_comp}
  \psi^{\rho'_{1},\rho'_{2}}_{a'}(p^{z},p^{xy};{\bf P}) &=&
    \int_{k_{z}>0}\! \frac{k^{xy}dk^{xy}k^{z}}{(2\pi)^{2}}
    G^{\rho'_{1},\rho'_{2}}_{a'}(p^{z},p^{xy};{\bf P})  \nonumber \\
  &&\hspace{.25in} \times
    V^{\rho'_{1},\rho'_{2}; \rho_{1},\rho_{2}}
    _{(k_{z}>0)\;a';a}(p^{z},p^{xy};k^{z},k^{xy})
     \psi^{\rho_{1},\rho_{2}}_{a}(k^{z},k^{xy};{\bf P}),
\end{eqnarray}
where the diagonal hermitian propagator is
\begin{eqnarray}
  && \int \frac{d\phi}{2\pi}
   \left[\chi^{\rho'_{1},\rho'_{2}}_{a'}({\bf p},{\bf P})\right]^{\dagger}
    g_{0}({\bf p};{\bf P}) \gamma^{0}_{1}\gamma^{0}_{2}
    \left[\chi^{\rho_{1},\rho_{2}}_{a}({\bf p},{\bf P})\right] = 
  G^{\rho'_{1},\rho'_{2}}(p^{z},p^{xy};{\bf P})
    \delta_{\rho'_{1},\rho_{1}} \delta_{\rho'_{2},\rho_{2}}
    \delta_{a',a} ,
\end{eqnarray}
with
\begin{eqnarray}
  G^{\rho'_{1},\rho'_{2}}(p^{z},p^{xy};{\bf P}) &\equiv&
       \left[ (\rho'_{1}+\rho'_{2}) \frac{E_{P}}{2} -
          \rho'_{1}\rho'_{2}(\epsilon_{1}+\epsilon_{2}) \right]^{-1} ,
\end{eqnarray}
and the hermitian potential operator is given by,
\begin{eqnarray}
\label{e:bVpotMat}
  && V^{\rho'_{1},\rho'_{2}; \rho_{1},\rho_{2}}_{a'; a}
    (p^{z},p^{xy};k^{z},k^{xy}) \equiv 
  \int\frac{d\phi_{p}d\phi_{k}}{(2\pi)^{2}}
    \left[\chi^{\rho'_{1},\rho'_{2}}_{a'}({\bf p},{\bf P})\right]^{\dagger}
    \!\!\gamma^{0}_{1}\gamma^{0}_{2} \hat{V}({\bf p},{\bf k};P)
    \chi^{\rho_{1},\rho_{2}}_{a}({\bf k},{\bf P})  
\end{eqnarray}
or
\begin{eqnarray}
\label{e:12bVpotMat}
 V^{\rho'_{1},\rho'_{2}; \rho_{1},\rho_{2}}_{(k_{z}>0)\;a'; a}
    (p^{z},p^{xy};k^{z},k^{xy}) &\equiv&
    \int\frac{d\phi_{p}d\phi_{k}}{(2\pi)^{2}}
    \left[\chi^{\rho'_{1},\rho'_{2}}_{a'}({\bf p},{\bf P})\right]^{\dagger}
  \nonumber\\ && \hspace{.1in} \times
    \gamma^{0}_{1}\gamma^{0}_{2} \left[ \hat{V}({\bf p},{\bf k};P) +
      \hat{V}({\bf p},-{\bf k};P) \eta_{12} {\cal P}_{\!\!ex} \right]
    \chi^{\rho_{1},\rho_{2}}_{a}({\bf k},{\bf P}) . 
\end{eqnarray}
A detailed partial-wave analysis of the potential is contained in appendix
\ref{app:Vpartial}.
Equation~(\ref{e:homoeq_comp}) or~(\ref{e:12homoeq_comp}) is solved for
$M_{J}=0$ at fixed values
of total momentum using the Malfliet-Tjon iteration procedure\cite{MalfTjon}
and numerical integration over $p^{xy}$ and $p^{z}$
(or radial and polar angle components).
Wave functions with polarization states $M_{J}=\pm 1$ are obtained from
the $M_{J}=0$ state by using the raising and lowering operator,
 $\psi^{M_{J}\pm 1} = \sqrt{(J+M_{J})(J-M_{J}+1)} J^{\pm} \psi^{M_{J}}$.

\section{Results for wave functions }
\label{sec:waves}

\begin{figure}
  \centerline{\epsfxsize=16cm  \epsffile{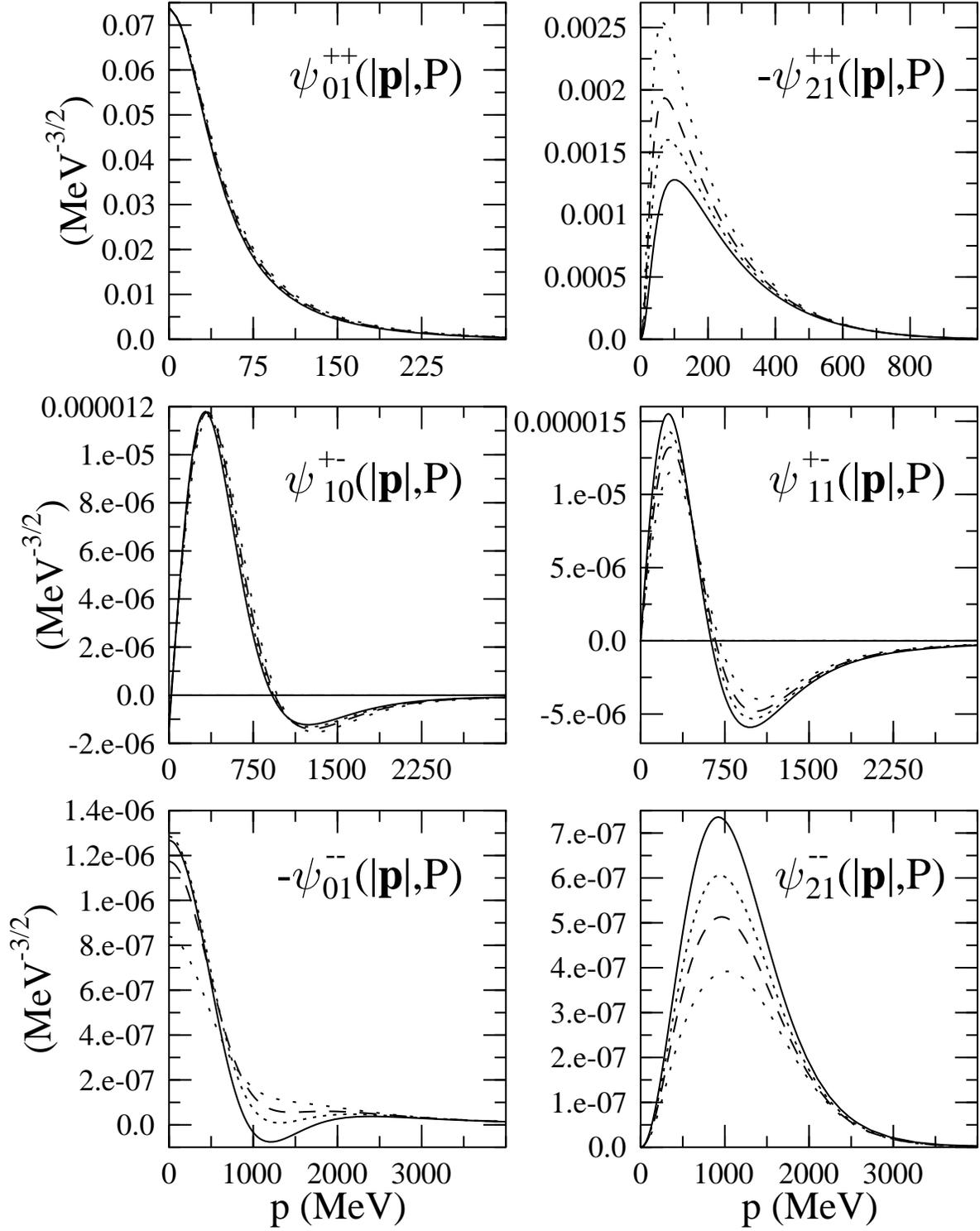} }
  \caption{Dependence of Breit-frame wave functions on total momentum.
           In this figure $J=1$ and $M_{J}=0$,
           $P\approx 0MeV$(solid), $P=698MeV$(dotted),
           $P=987MeV$(dashed), and $P=1395MeV$(long dots).
           }\label{f:bvwfn}
\end{figure}

Our wave functions vary as a function of both the magnitude and polar angle
of relative momentum.
To show how the wave functions change with total momentum,
we project them onto standard LSJ basis functions
and integrate out dependence on the polar angles,
\begin{equation}
\label{e:bvwfn}
  \psi^{\rho_{1}\rho_{2}}_{lSJM_{J}}(|{\bf p}|,{\bf P}) \equiv \int d\Omega_{p}
    \left[ {\cal Y}_{lSJM_{J}}(\hat{p}) \right]^{\dagger}
     \psi^{M_{J}}({\bf p},{\bf P}) ,
\end{equation}
where the LSJ basis functions are given by,
\begin{eqnarray}
  {\cal Y}^{M_{J}}_{lSJ}(\hat{\bf p}) = \sum^{+1}_{m_{S}=-1} &&
      \langle l,S,m_{l},m_{S} | l,S,J,M_{J} \rangle  
        Y_{l,m_{l}}(\theta,\phi) \: |S,m_{S}\rangle ,
\end{eqnarray}
where $l$ and $S$ are the relative-orbital and spin angular momenta,
${\bf j}={\bf l}+{\bf S}$ and ${\bf l}={\bf r}\times{\bf p}$.
These wave function components are shown in Fig.~\ref{f:bvwfn} for
total momentum corresponding to ${\bf P}^{2}=$.0025, 12.5, 25 and 50$fm^{-2}$.
A `probability' for the wave function projections as defined by
\begin{equation}
  P^{\rho_{1}\rho_{2}}_{lS} \equiv \frac{M}{E_{P}} \int p^{2}dp
    \left| \psi^{\rho_{1}\rho_{2}}_{lS}(|{\bf p}|,{\bf P}) \right|^{2}
\end{equation}
is shown in Table~\ref{t:psiProjProbs} based on
the wave function normalization of Eq.~(\ref{e:psiNorm}).
%
\begin{table} 
\begin{center} \begin{tabular}{cllll} 
 & \multicolumn{4}{c}{${\bf P}^{2}$}\\ 
$\rho_{1}\rho_{2}lS$ & 0.0025$fm^{-2}$	& 12.5$fm^{-2}$
 			& 25$fm^{-2}$	& 50$fm^{-2}$ \\
\tableline
$++$01 & 0.9499 & 0.9433 & 0.9363 & 0.9226 \\
$++$21 & 4.97$\times10^{-2}$ & 5.59$\times10^{-2}$
       & 6.18$\times10^{-2}$ & 7.22$\times10^{-2}$ \\
$+-$10 & 3.86$\times10^{-5}$ & 4.00$\times10^{-5}$
       & 4.09$\times10^{-5}$ & 4.18$\times10^{-5}$ \\
$+-$11 & 1.46$\times10^{-4}$ & 1.19$\times10^{-4}$
       & 9.80$\times10^{-5}$ & 6.86$\times10^{-5}$ \\
$--$01 & 2.70$\times10^{-7}$ & 3.07$\times10^{-7}$
       & 3.20$\times10^{-7}$ & 2.82$\times10^{-7}$ \\
$--$21 & 2.49$\times10^{-6}$ & 1.72$\times10^{-6}$
       & 1.27$\times10^{-6}$ & 8.01$\times10^{-7}$ \\
$P$(++) & 0.99962 & 0.99921 & 0.99815 & 0.99475 \\
\end{tabular} \end{center}
\caption{Wave function projection probabilities ($J=1$ and $M_{J}=0$).
 See text. $P(++) \equiv P($++$01) + P($++$21)$.
\label{t:psiProjProbs} }
\end{table}

\section{Concluding remarks}

The theory of relativistic bound states is formulated in three
dimensions by use of an instant reduction of the Bethe-Salpeter
formalism.  This formalism is applicable to the analysis of
elastic form factors and it is consistent with current
conservation and four-momentum conservation.  Elastic form
factors involve matrix elements which are calculated in the
Breit frame, thus requiring wave functions for the initial and
final states that have been boosted to momentum ${\bf P} = \pm
{1 \over 2} {\bf q}$.  These wave functions are calculated in
this paper for the case of the deuteron.
The form factors for the deuteron will be
the subject of another paper.

    The main issues addressed in this work are the boost of the
interaction that is required and the solution of the
quasipotential wave equation in frames where the total momentum
is nonzero.  We have shown that in general the boosted
interaction is defined as the solution of a four-dimensional
equation and that equation has been solved for the special case
of a scalar, separable interaction.  The main effect of the
boost is to renormalize the dominant matrix elements of the
interaction as three-momentum varies, although there is also in
general a change in the Dirac structure of the interaction.
An approximation which captures the renormalization effect is
used for the more complicated one-boson exchange interaction.
For momenta up to ${\bf P}^{2}$ = 50 fm$^{-2}$, the renormalization
of the interaction is modest in the case of the deuteron,
varying linearly with ${\bf P}^{2}$ and
amounting to about a 10\% reduction of the one-boson exchange
potential at ${\bf P}^{2}$ = 50 fm$^{-2}$.
A complete solution of the boost of the meson-exchange
interaction is left
as an unsolved problem.  However, the separable potential analysis
suggests that a perturbative expansion of Eq.~(\ref{e:boost}) may
provide accurate results for the boost.

    The solutions for the quasipotential wave functions have
been developed and the results show that there are modest
variations as the deuteron is boosted to momenta up to ${\bf P}^{2} =
$50 fm$^{-2}$, corresponding to ${\bf q}^{2}$ = 200 fm$^{-2}$.
In a future article, they will be applied to the calculation
of elastic form factors for the deuteron.


   Support for this work by the U.S. Department of
Energy under grants DE-FG02-93ER-40762 and DE-AC05-84ER40150 is gratefully
acknowledged.

\appendix

\section{Basis functions and operators}
\label{app:spinors}


The hermitian plane-wave spinors for particle $i$ and denoted by
$u_{i}^{\rho_{i}}(\rho_{i}{\bf k})$ are
\begin{eqnarray}
\label{e:hpwspinors}
  u^{+}_{i}(+{\bf k}_{i}) = N \left( \begin{array}{c} 
          \mbox{\large 1} \\
          \frac{\textstyle \bbox{\sigma}_{i}\cdot {\bf k}_{i}}
               {\textstyle \epsilon_{i} + m}
          \end{array} \right) , \nonumber \\
  u^{-}_{i}(-{\bf k}_{i}) = N \left( \begin{array}{c}
           \frac{\textstyle -\bbox{\sigma}_{i}\cdot {\bf k}_{i}}
                {\textstyle \epsilon_{i} + m} \\
           \frac{\textstyle}{\textstyle}  \mbox{\large 1} \end{array} \right),
\end{eqnarray}
with $N = \left( \frac{\epsilon_{i}+m}{2\epsilon_{i}} \right)^{1/2}$
and $\epsilon_{i} = \left( m^{2} +{\bf k}_{i}^{2} \right)^{1/2}$.
It will be convenient later to use spinors with zero momentum
which are related to the plane-wave spinors by,
\begin{eqnarray}
\label{e:pwulbasistrans}
  u^{\rho_{i}}_{i}(\rho_{i}{\bf k}_{i}) = N
   ( 1 + \rho_{i} \frac{\bbox{\alpha_{i}}\cdot{\bf k}_{i} }{ \epsilon_{i}+m } )
   u^{\rho_{i}}_{i}({\bf 0}) .
\end{eqnarray}

Symmetry and Dirac operators acting on the basis functions
will produce linear combinations (independent of $\phi$) of the
basis functions.  Some useful
examples of this follow.  The action of the operators on the
Dirac and spin parts of the basis functions will be considered
separately.  Consider first the symmetry operators of
Eqs.~(\ref{e:parityop}) -- (\ref{e:timeparityop}),
\begin{eqnarray}
  {\cal P}      u^{\rho_{i}}_{i}(\rho_{i}{\bf k}_{i}) &=&
      \rho_{i}  u^{\rho_{i}}_{i}(\rho_{i}{\bf k}_{i}) {\cal P}_{s} , \\
  {\cal P}_{12} u^{\rho_{i}}_{i}(\rho_{i}{\bf k}_{i}) &=&
                u^{\rho_{i}}_{j}(\rho_{i}{\bf k}_{j}) {\cal P}_{12} \;
                                                          (j\neq i) , \\
  {\cal T}      u^{\rho_{i}}_{i}(\rho_{i}{\bf k}_{i}) &=&
                u^{\rho_{i}}_{i}(\rho_{i}{\bf k}_{i}) {\cal T} ,
\end{eqnarray}
and
\begin{eqnarray}
  {\cal P}_{s} {\cal Y}^{M_{J}}_{s_{1},s_{2}}(\phi) &=&
    (-1)^{M_{J}-s_{1}-s_{2}}  {\cal Y}^{M_{J}}_{s_{1},s_{2}}(\phi) , \\
  {\cal P}_{12} {\cal Y}^{M_{J}}_{s_{1},s_{2}}(\phi) &=&
    (-1)^{M_{J}-s_{1}-s_{2}}  {\cal Y}^{M_{J}}_{s_{2},s_{1}}(\phi) , \\
  {\cal T} {\cal Y}^{M_{J}}_{s_{1},s_{2}}(\phi) &=&
    (-1)^{M_{J}} {\cal Y}^{-M_{J}}_{-s_{1},-s_{2}}(\phi) .
\end{eqnarray}
Similar relations for the ${\cal Y}^{M_{J}}_{\alpha}(\phi)$
of Eq.~(\ref{e:Yofphi}) $(\alpha\in \{e1,o1,e0,o0\})$
are easily derived from the above equations.  In particular,
for the basis $\chi^{\rho_{1},\rho_{2}}_{M_{J},\alpha}({\bf k},{\bf P}) =
 u^{\rho_{1}}_{1}(\rho_{1}{\bf k}_{1}) u^{\rho_{2}}_{2}(\rho_{2}{\bf k}_{2})
 {\cal Y}^{M_{J}}_{\alpha}(\phi)$,
\begin{eqnarray}
  {\cal P} {\cal T} \chi^{\rho_{1},\rho_{2}}_{\alpha,M_{J}}({\bf k},{\bf P})
  &=& \left\{ \begin{array}{cc}
  +\rho_{1}\rho_{2} \chi^{\rho_{1},\rho_{2}}_{\alpha,-M_{J}}
                    ({\bf k},{\bf P}) & \alpha=o1,e0 \\
  -\rho_{1}\rho_{2} \chi^{\rho_{1},\rho_{2}}_{\alpha,-M_{J}}
                    ({\bf k},{\bf P}) & \alpha=e1,o0 \\
  \end{array} \right. . \nonumber\\
\end{eqnarray}
Thus these basis functions have ${\cal P} {\cal T} = \pm 1$ if $M_{J}=0$.
The condition ${\cal P} {\cal T} = \pm 1$ is equivalent to
$Parity = {\cal P} = \pm (-1)^{J+1}$ for the standard LSJ basis functions
in the center-of-mass frame.

Consider next Dirac operators such as $\gamma^{0}_{i}$,
$\gamma_{i}\cdot k_{i}$, $\gamma_{1}\cdot\gamma_{2}$, etc.,
acting on the simple basis functions
given by $\chi^{\rho_{1},\rho_{2}}_{s_{1},s_{2}}({\bf k},{\bf P}) =
 u^{\rho_{1}}_{1}({\bf 0}) u^{\rho_{2}}_{2}({\bf 0})
 {\cal Y}^{M_{J}}_{s_{1},s_{2}}(\phi)$.
Their effect on the zero-momentum spinors is particularly simple,
\begin{eqnarray}
\label{e:gammau}
  \gamma^{0}_{i} u^{\rho_{i}}_{i}({\bf 0})
    &=& \rho_{i} u^{\rho_{i}}_{i}({\bf 0}) , \\
\label{e:gamma5u}
  \gamma^{5}_{i} u^{\rho_{i}}_{i}({\bf 0})
    &=&         u^{-\rho_{i}}_{i}({\bf 0}) .
\end{eqnarray}
Moreover, spin operators acting on the basis functions
produce linear combinations as follows,
\begin{eqnarray}
\label{e:sigmaY}
  \bbox{\sigma}_{1}\cdot \bbox{\sigma}_{2} {\cal Y}^{M}_{s_{1},s_{2}} &=&
     4 s_{1} s_{2} {\cal Y}^{M}_{s_{1},s_{2}} +
      2 \delta_{s_{1},-s_{2}} {\cal Y}^{M}_{-s_{1},-s_{2}} , \\
\label{e:sigma1Y}
  \bbox{\sigma}_{1}\cdot {\bf k}_{1} {\cal Y}^{M}_{s_{1},s_{2}} &=&
     2 s_{1} k^{z}_{1} {\cal Y}^{M}_{s_{1},s_{2}} +
            k^{xy}_{1} {\cal Y}^{M}_{-s_{1},s_{2}} , \\
\label{e:sigma2Y}
  \bbox{\sigma}_{2}\cdot {\bf k}_{2} {\cal Y}^{M}_{s_{1},s_{2}} &=&
     2 s_{2} k^{z}_{2} {\cal Y}^{M}_{s_{1},s_{2}} +
            k^{xy}_{2} {\cal Y}^{M}_{s_{1},-s_{2}} ,
\end{eqnarray}
where $k^{xy}_{1} = k^{xy} = -k^{xy}_{2}$.
For the analysis of potential operators, it is useful to note,
\begin{eqnarray}
  && \bbox{\sigma}_{1}\cdot {\bf q}_{1}
     {\cal Y}^{M}_{s_{1},s_{2}}(\phi_{k}) =
     2 s_{1} q^{z}_{1} {\cal Y}^{M}_{s_{1},s_{2}}(\phi_{k})
  + (p^{xy}_{1} e^{2s_{1}i\phi_{d}} - k^{xy}_{1})
            {\cal Y}^{M}_{-s_{1},s_{2}}(\phi_{k}) , \\
  && \bbox{\sigma}_{2}\cdot {\bf q}_{2}
     {\cal Y}^{M}_{s_{1},s_{2}}(\phi_{k}) =
     2 s_{2} q^{z}_{2} {\cal Y}^{M}_{s_{1},s_{2}}(\phi_{k})
  + (p^{xy}_{2} e^{2s_{2}i\phi_{d}} - k^{xy}_{2})
            {\cal Y}^{M}_{s_{1},-s_{2}}(\phi_{k}) ,
\label{e:sigmapk2}
\end{eqnarray}
where ${\bf q}_{i}\equiv{\bf p}_{i}-{\bf k}_{i}$ and
$\phi_{d}\equiv\phi_{p}-\phi_{k}$.

Similar relations can be derived for the same Dirac operators acting on the
sixteen $u^{\rho_{1}}_{1}({\bf 0}) u^{\rho_{2}}_{2}({\bf 0})
 {\cal Y}^{M_{J}}_{\alpha}(\phi)$ basis functions.
This can be accomplished by using the
above equations and defining a basis transform between the two bases.
Note that some of the above operators, such as
$\gamma^{5}_{i}$, do not commute with ${\cal P}{\cal T}$.

\section{Partial-Wave Potential Analysis}
\label{app:Vpartial}

In this appendix we sketch the partial wave analysis of the
one-boson-exchange potential.
This is most readily accomplished using the
$\chi^{\rho_{1}\rho_{2}}_{s_{1},s_{2}} =
 u^{\rho_{1}}_{1}({\bf 0}) u^{\rho_{2}}_{2}({\bf 0})
 {\cal Y}^{M_{J}}_{s_{1},s_{2}}(\phi)$ basis functions.
The potential in this basis can easily be transformed to
plane-wave or PT bases.
The operator form of the potential is generated from the Feynman rules
for meson propagators,
\begin{eqnarray}
   \Delta(q) &=& \frac{1}{q^{2} - \mu^{2} +i\eta} , \\
   \Delta_{\alpha,\beta}(q) &=& \left[ -g_{\alpha,\beta} +
      \frac{q_{\alpha}q_{\beta}}{\mu^{2}} \right] \Delta(q) ,
\end{eqnarray}
and meson-nucleon vertices,
\begin{eqnarray}
   \Lambda_{S}(q) &=& -1 , \\
  \Lambda_{P}(q) &=& i \gamma\cdot q \gamma^{5} , \\
   \Lambda_{V}^{\alpha}(q) &=& (\gamma^{\alpha} + \frac{f}{2m}
       i \sigma^{\alpha,\nu} q_{\nu} ) ,
\end{eqnarray}
where the exchanged four momentum is $q = p_{1}-k_{1} = p-k = k_{2}-p_{2}$.
The exchange of a single meson is given by $\Lambda(q)\Delta(q)\Lambda(-q)$
(or $\Lambda^{\alpha}(q)\Delta_{\alpha,\beta}(q)\Lambda^{\beta}(-q)$ for
vector mesons) with a factor of $\tau_{1}\cdot\tau_{2}$ added to the
exchange of isovector mesons and a coupling constant $g$ and form
factor $F(q^{2})$ attached to each meson-nucleon vertex.
Note that for the isoscalar deuteron,
${\bf \tau}_{1}\cdot{\bf \tau}_{2} \rightarrow -3$.
The Bonn model form factors are,
\begin{eqnarray}
\label{e:BonnFF}
  F_{i}(q^{2}) &=& \frac{\Lambda^{2}_{i}-\mu^{2}_{i} }{
      \Lambda^{2}_{i} -q^{2} } .
\end{eqnarray}
Thus the full one-boson-exchange potential is,
\begin{eqnarray}
  \hat{V}(p,k) &=& \sum_{i} g^{2}_{i} t_{i} \Delta_{i}(q) F^{2}_{i}(q^{2})
    \hat{V}_{i}(q) ,
\end{eqnarray}
where $i$ is summed over the six mesons of Table~\ref{t:mesonparams},
\begin{equation}
   t_{i} \equiv \left\{ \begin{array}{ccc}
    \tau_{1}\cdot\tau_{2} & if & i = isovector\;meson \\
                        1 & if & i = isoscalar\;meson \end{array} \right\} ,
\end{equation}
and
\begin{eqnarray}
   \hat{V}_{S}(q)  &\equiv& \Lambda_{S}(q)\Lambda_{S}(-q) , \\
   \hat{V}_{P}(q) &\equiv& -\Lambda_{P}(q)\Lambda_{P}(-q) , \\
   \hat{V}_{V}(q) &\equiv& \Lambda^{\alpha}_{V}(q)
        \left[ -g_{\alpha,\beta} + \frac{q_{\alpha}q_{\beta}}{\mu^{2}} \right]
       \Lambda^{\beta}_{V}(-q) .
\end{eqnarray}
Using Eqs.~(\ref{e:gammau}) to (\ref{e:sigmapk2}) it can be seen that
\begin{eqnarray}
\label{e:VnumBVbasis}
  && \gamma^{0}_{1}\gamma^{0}_{2}\hat{V}_{i}(q) \chi^{\rho_{1},\rho_{2}}
     _{s_{1},s_{2},M_{J}}({\bf k},{\bf P})
     = \chi^{\rho''_{1},\rho''_{2}}_{s''_{1},s''_{2},M_{J}}({\bf k},{\bf P})
   [V_{i}(p^{z},p^{xy},k^{z},k^{xy})]^{\rho''_{1},\rho''_{2},\rho_{1},\rho_{2}}
     _{s''_{1},s''_{2},s_{1},s_{2},M_{J};n} e^{i n \phi_{d}} ,
\end{eqnarray}
where $i=S,P,V$, $n=0,\pm 1,\pm 2$ and $\phi_{d}\equiv\phi_{p}-\phi_{k}$
with implicit sum over double prime variables.
The exact form of $[V_{i}]$ is easily derived using a symbolic manipulation
program such as Mathematica.
For $\hat{V}_{V}$, note $\cos(\phi_{d})=(e^{i\phi_{d}}-e^{-i\phi_{d}})/2$.
Next, the scalar part of the potential, which is a function of $\phi_{d}$
but not of $\phi_{p}$ or $\phi_{k}$ alone, can be integrated,
\begin{eqnarray}
  && \int \frac{d\phi_{p}d\phi_{k}}{(2\pi)^{2}} \left[
    \chi^{\rho'_{1},\rho'_{2}}_{s'_{1},s'_{2},M'_{J}}(\phi_{p})
       \right]^{\dagger} \Delta F^{2} e^{i n \phi_{d}}
    \left[\chi^{\rho''_{1},\rho''_{2}}_{s''_{1},s''_{2},M_{J}}(\phi_{k})\right]
  = \delta_{\rho'_{1},\rho''_{1}} \delta_{\rho'_{2},\rho''_{2}}
      \delta_{s'_{1},s''_{1}} \delta_{s'_{2},s''_{2}} \delta_{M'_{J},M_{J}}
      I_{F\Delta F}^{n-l} ,
  \nonumber\\
\end{eqnarray}
where $l\equiv M_{J}-s'_{1}-s'_{2}$ and
\begin{equation}
\label{e:scalarMEC}
  I_{F\Delta F}^{n} \equiv \int \frac{d\phi_{d}}{2\pi}
     \Delta(q) F^{2}(q^{2}) e^{i n \phi_{d}} .
\end{equation}

%
\begin{table}
\begin{center} \begin{tabular}{ccccccc}
\multicolumn{7}{c}{Bonn B potential parameters} \\
\multicolumn{7}{c}{(energy independent, Thompson propagator)} \\
\tableline
 & $\pi$ & $\eta$ & $\rho$ & $\omega$ & $\delta$ & $\sigma$  \\
\tableline
 & PV-IV & PV-IS & V-IV & V-IS & S-IV & S-IS \\
        $\mu$  & .13803 & .5488 & .769     & .7826 & .983   & .550  \\
   $\Lambda$   & 1.2    & 1.5   & 1.3      & 1.5   & 1.5    & 2.0    \\
  $g^{2}/4\pi$ & 14.6   & 5.0   & .95      & 20.0  & 3.1155 & 8.0769  \\
  $f$	       &	&	& 6.1      & 0     &	    &          \\
\multicolumn{6}{l}{$g^{2}_{\sigma}/4\pi$ with negative energy sectors}
& 8.5503 \\
\end{tabular} \end{center}
\caption{Meson Parameters of the Bonn B model  \protect\cite{BonnABC},
 and modification to
 $g^{2}_{\sigma}/4\pi$ when negative energy sectors are included.
 $M$ and $\Lambda$  in  $GeV$. Nucleon mass is $m=.938926GeV$,
 deuteron binding energy is 2.224644MeV. Vector propagator is
 approximated by $\Delta_{\alpha,\beta}(q)\rightarrow
 -g_{\alpha,\beta} \Delta(q)$.\label{t:mesonparams}}
\end{table}
%
Combining the necessary factors, the partial-wave potential is,
\begin{eqnarray}
  && [V(p^{z},p^{xy},k^{z},k^{xy})]^{\rho'_{1},\rho'_{2},\rho_{1},
      \rho_{2}}_{s'_{1},s'_{2},s_{1},s_{2},M_{J}} =
  \sum_{i} g^{2}_{i} t_{i} I_{F\Delta F}^{n-l}
     [V_{i}(p^{z},p^{xy},k^{z},k^{xy})]^{\rho''_{1},\rho''_{2},
     \rho_{1},\rho_{2}}_{s''_{1},s''_{2},s_{1},s_{2},M_{J};n} .
\end{eqnarray}
This partial-wave potential based on zero-momentum spinors
can be transformed to a plane-wave
basis using the basis transform implied by Eqs.~(\ref{e:pwulbasistrans}),
(\ref{e:gamma5u}), and~(\ref{e:sigma1Y}--\ref{e:sigma2Y}),

Finally, the scalar $\phi_{d}$ integral of Eq.~(\ref{e:scalarMEC}) must be
evaluated.  This can be accomplished analytically using $z=e^{i\phi}$
and contour integration around the unit circle in the complex plane.
We first express $\Delta F^{2}$ as a sum of simple denominators times
$\phi_{d}$ independent coefficients using a partial fraction expansion.
The integral of a simple meson denominator is given by,
\begin{eqnarray}
\label{e:simpleIphi}
  I_{\mu}^{n} &\equiv& \int\frac{d\phi_{d}}{2\pi} \frac{e^{i n \phi_{d}}
    }{ a_{\mu} - b \cos(\phi_{d}) } = -\int\frac{d\phi_{d}}{2\pi} \Delta(q)
    e^{i n \phi_{d}} = \frac{z_{\mu}^{n}}{S_{\mu}} , 
\end{eqnarray}
where $a_{\mu} \equiv \mu^{2}_{i} -(q(\phi_{d}=0))^{2}$,
 $b \equiv 2 p_{xy}k_{xy}$
(thus $-\Delta^{-1}(q) = \mu^{2}-q^{2} = a_{\mu} - b \cos(\phi_{d})$),
and $a_{\mu}>b$, $z_{\mu}\equiv(a-S)/b$,
 $S_{\mu}\equiv\sqrt{a^{2}-b^{2}}$.
The denominators of the Bonn form factors, Eq.~(\ref{e:BonnFF}),
take the same form with $\mu$ replaced by $\Lambda$; thus
$I_{\Lambda}\equiv I_{\mu}(\mu\rightarrow\Lambda)$.
Using the partial fraction expansion for Bonn form factors
and the simple integrals above,
\begin{equation}
  I_{F\Delta F}^{n} = - I_{\mu}^{n} + I_{\Lambda}^{n} - (\Lambda^{2}-\mu^{2})
      \frac{\partial}{\partial a_{\Lambda}} I_{\Lambda}^{n} .
\end{equation}
This integral can be numerically checked using the following simple,
well known, technique.
We can write,
\begin{eqnarray}
   \Delta(q) F^{2}(q^{2}) &=& \Delta(q) +
      \left( F^{2}(\mu^{2})-F^{2}(q^{2}) \right) \Delta(q) ,
\end{eqnarray}
for any form factors with $F(\mu^{2})=1$.
The first term is integrated using  Eq.~(\ref{e:simpleIphi}), while
the second term is evaluated numerically.  Note the second term
is non-singular even at $q^{2}=\mu^{2}$,
which can not occur with instant constraints,
but can occur if different constraints are used on the left and right of V.


\end{document}